\documentclass[aip,amsmath,amssymb,reprint]{revtex4-1}

\usepackage{graphicx}
\usepackage{dcolumn}
\usepackage{bm}
\usepackage[utf8]{inputenc}
\usepackage[T1]{fontenc}
\usepackage{mathptmx}
\usepackage{etoolbox}
\usepackage[separate-uncertainty=true]{siunitx}

\makeatletter
\def\@email#1#2{
 \endgroup
 \patchcmd{\titleblock@produce}
  {\frontmatter@RRAPformat}
  {\frontmatter@RRAPformat{\produce@RRAP{*#1\href{mailto:#2}{#2}}}\frontmatter@RRAPformat}
  {}{}
}
\makeatother
\begin{document}

\preprint{AIP/123-QED}

\title[Polarizing antiresonant hollow-core fiber]{Polarizing Antiresonant Hollow-Core Fiber}

\author{Yuxi Wang}
 \affiliation{School of Electrical and Electronic Engineering, Nanyang Technological University, 50 Nanyang Avenue, 639798, Singapore}
 
\author{Charu Goel}
 \affiliation{Temasek Laboratories, Nanyang Technological University, 50 Nanyang Avenue, 639798, Singapore}

\author{Wonkeun Chang}
 \email{wonkeun.chang@ntu.edu.sg}
 \affiliation{School of Electrical and Electronic Engineering, Nanyang Technological University, 50 Nanyang Avenue, 639798, Singapore}

\date{\today}

\begin{abstract}
Achieving robust single-polarization guidance in hollow-core fibers has remained a longstanding challenge, limiting their integration into precision photonic systems. Here, we report the first experimental realization of a low-loss, polarization filtering antiresonant hollow-core fiber (AR-HCF). Conventional AR-HCFs inherently support degenerate orthogonal polarization modes, making them vulnerable to polarization drift under environmental perturbations. Our dual-ring fiber design introduces polarization-selective resonant coupling to lossy cladding modes, enabling strong polarization filtering without compromising transmission efficiency. The fiber achieves a polarization extinction ratio exceeding \SI{21}{\deci\bel} and a propagation loss as low as \SI{0.15}{\deci\bel\per\meter} over a \SI{10}{\meter} fiber length. The design is scalable across wavelength bands and maintains polarization discrimination under mechanical bending, making it highly suitable for applications in fiber-based gyroscopes, quantum optics, and polarization-sensitive nonlinear interactions. This work represents a significant step toward monolithic, polarization-selective hollow-core fiber systems.
\end{abstract}

\maketitle

\section{\label{intro}Introduction}

Recent advances in hollow-core fibers are paving the way for breakthroughs in optical telecommunications systems \cite{1}, sensor technologies \cite{2,3,4}, and high-power laser beam delivery \cite{5}. Light guidance through the hollow core provides benefits such as low loss, low latency, low nonlinearity, high power-damage threshold, and low thermal and electromagnetic sensitivity \cite{6}. Among the variety of hollow-core fibers, tubular-cladding antiresonant hollow-core fibers (AR-HCFs) have received significant attention due to their ultra-low transmission loss, which can even be lower than the loss of telecommunication fibers \cite{1}, broadband guidance over octave-wide bandwidths \cite{7}, effective single-mode operation with a large fundamental mode (FM) area \cite{8}, and relatively easy fabrication \cite{9}.

An AR-HCF intrinsically supports multiple core modes. However, higher-order modes can be effectively suppressed along the fiber length by introducing a high differential loss through carefully designed cladding \cite{8}. Within the FM, light propagates in two degenerate orthogonally polarized modes \cite{10}. While a loosely wound AR-HCF can exhibit excellent polarization purity under static conditions \cite{11}, environmental perturbations such as bending, twisting, and temperature changes may induce polarization mode coupling \cite{12}. This can be detrimental. For instance, noise arising from polarization drift may impact the precision of interferometric sensors like gyroscopes, which are widely used in the navigation and guidance systems of airplanes, spacecraft, and ships \cite{13}. Therefore, while standard AR-HCFs can significantly enhance the performance of conventional fiber gyroscopes, a free-space coupled polarizer in the setup is still necessary for additional polarization control \cite{3}. This makes the system susceptible to atmospheric turbulence, optical system aberration, and fiber positioning errors, compromising overall robustness. Replacing the standard AR-HCF with a single-polarization AR-HCF may yield a more reliable monolithic gyroscope. Likewise, a single-polarization AR-HCF can enhance the performance of bending or strain sensors, as mechanical stresses may otherwise induce non-degeneracy of the two orthogonally polarized fundamental core modes in a standard AR-HCF \cite{14}. Single-polarization AR-HCFs also find applications in systems utilizing light-matter interactions in gas-filled AR-HCFs, such as atom optics \cite{14}, polarized supercontinuum generation \cite{15}, and gas fiber lasers \cite{16}. In addition, quantum interference experiments such as the Hong–Ou–Mandel effect and boson sampling require photons with indistinguishable spatial-temporal modes and polarization states to ensure perfect interference \cite{17,18}. Single-polarization AR-HCF offers an effective all-fiber platform by permitting only a specific linearly polarized mode with high polarization purity, enhancing quantum interference fidelity while reducing errors caused by misalignment \cite{19}. They can also be used as functional fiber elements in polarization-sensitive spectroscopy and linearly polarized fiber laser cavities \cite{20,21,22,23}.

Preventing polarization mode coupling and ensuring single polarization guidance in solid-core fibers is relatively straightforward, achieved by introducing stress-induced birefringence or utilizing elliptical cores \cite{24,25,26}. However, in AR-HCFs, stress-induced birefringence is not feasible due to light confinement in the hollow core. In a standard AR-HCF, the two orthogonal polarizations of the FM are perfectly degenerate with a birefringence less than $10^{-8}$ \cite{11,27}. A weakly birefringent AR-HCF with a maximum birefringence of $9.1\times10^{-5}$ can be designed by exploiting polarization-dependent interactions of the FM with dielectric cladding modes in a bi-thickness nested structure \cite{27}. However, this scheme cannot suppress any light in the unwanted polarization and hence, cannot ensure single polarization guidance. Moreover, birefringent fibers suffer from polarization mode dispersion, which may distort short optical pulses \cite{28}. To achieve high birefringence in hollow-core polarization-maintaining fibers, a small core size is generally required \cite{27,29}, which limits the fiber's applications requiring large mode area. Therefore, for robust single polarization guidance, the AR-HCF needs to offer a finite polarization-dependent loss to suppress light in the unwanted polarization state \cite{10,29,30,31,32,33,34,35}.

Zang et al.~reported a bi-thickness single-ring AR-HCF with an impressive polarization suppression exceeding \SI{25}{\deci\bel}, achieved in only a \SI{6}{\centi\meter}-long fiber. However, this fiber suffers from a large loss for the pass polarization, restricting its use as a discrete polarizing device \cite{32}. Achieving a substantial polarization-dependent loss while maintaining low propagation loss in the pass polarization necessitates an asymmetric fiber geometry with two sets of independent structural parameters. Several AR-HCF designs have been theoretically explored in this direction \cite{10,29,30}. However, most of them require stringent control of structural parameters or involve geometries posing significant fabrication challenges \cite{10,29}. Hence, all these design proposals remain unverified experimentally until now.

In this work, we employ the bi-thickness dual-ring AR-HCF framework to demonstrate a low-loss single polarization mode hollow-core fiber \cite{30,36,37}. This is accomplished by introducing an asymmetry in the inner cladding layer of the dual-ring AR-HCF structure to suppress one of the FM orthogonal polarizations, while supporting the confinement of the pass polarization by a circularly symmetric outer cladding layer.

\section{\label{design}Design Principles and Fiber Geometry}

\begin{figure*}
\includegraphics{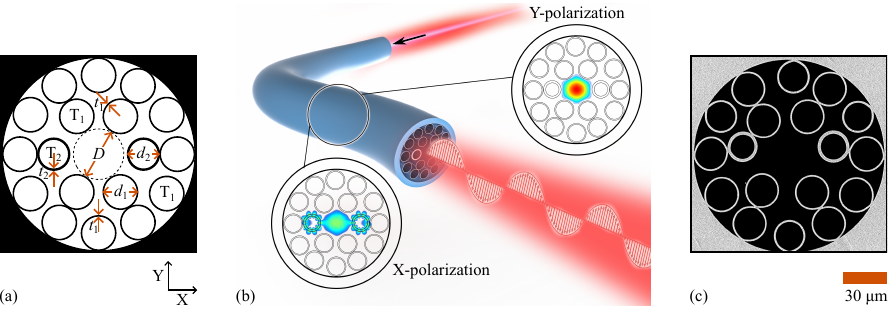}
\caption{\label{fig1}(a) Schematic of an idealized single-polarization double-ring hollow-core fiber (SP-DRF) cross-section. (b) Illustration of the SP-DRF operating principle. (c) Scanning electron microscopic image of the SP-DRF cross-section.}
\end{figure*}

Figure \ref{fig1}a shows the cross-section of an idealized single-polarization dual-ring hollow-core fiber (SP-DRF). It consists of two layers of cladding tubes surrounding the hollow core \cite{36}. We label the two inner cladding tubes along the X-axis T\textsubscript{2} and all the other inner and outer layer cladding tubes as T\textsubscript{1}. We define the core diameter of the SP-DRF as $D$, the inner diameter of the cladding tubes T\textsubscript{1} and T\textsubscript{2} as $d_1$ and $d_2$, and their wall thicknesses as $t_1$ and $t_2$, respectively.

An AR-HCF guides light in the hollow core by inhibiting the coupling of the fundamental core modes to the cladding modes \cite{38,39,40}. Resonant coupling between the modes in the hollow core and dielectric cladding walls near cutoff wavelengths of dielectric modes creates the typical AR-HCF transmission spectrum, which consists of periodic low- and high-loss spectral bands defined as the antiresonant and resonant bands, respectively. The resonant band edges are located at wavelengths given by:
\begin{equation}
\lambda_m=\frac{2t}{m}\sqrt{n_g^2-1}\textrm{,}
\label{eq1}
\end{equation}
where $t$ is the wall thickness of cladding tubes, $n_g$ is the refractive index of cladding glass, and $m$ is an integer denoting the resonance order. Several sharp loss peaks are observed in the wavelength region shorter than $\lambda_m$, wherein each loss peak corresponds to resonant coupling of FM to discrete cladding modes. In a symmetric AR-HCF, this resonant mode coupling is not polarization-dependent, since both orthogonal linear polarizations of the FM are nearly degenerate. The asymmetry in the inner cladding layer of the SP-DRF breaks this degeneracy, making the resonant coupling polarization-dependent. The wall thickness $t_2$ is chosen such that the operating wavelength lies at the short wavelength edge of a resonant band of T\textsubscript{2} tubes. The birefringence introduced by the structural asymmetry in the SP-DRF ensures that only the X-polarization of the FM couples to a dielectric cladding mode at this wavelength, while the Y-polarization is well-guided. The wall thickness $t_1$ is chosen such that the operation wavelength lies in the antiresonant band of T\textsubscript{1} tubes, to support the confinement of Y-polarization in the hollow core. Hence, the SP-DRF can simultaneously achieve low-loss propagation for the FM Y-polarization and high-loss propagation for the FM X-polarization, as illustrated in Figure \ref{fig1}b.

Figure \ref{fig1}c shows a scanning electron microscopic image of the SP-DRF cross-section, which was fabricated using the stack-and-draw method \cite{37}. The fiber has $D=$\SI{36}{\micro\meter}, $t_1=$ \num{1.30\pm0.02} \si{\micro\meter}, and $d_1/D=$ \num{0.71\pm0.01}. This value of $d_1/D$ can efficiently suppress higher-order core modes, leading to effective single-mode guidance in the SP-DRF \cite{8}. Due to fabrication challenges, $t_2$ and $d_2$ are slightly different in the two T\textsubscript{2} tubes. One has $t_2=$ \num{2.41\pm0.01} \si{\micro\meter} and $d_2/D=$ \num{0.65\pm0.01}, while the other has $t_2=$ \num{2.36\pm0.01} \si{\micro\meter} and $d_2/D=$ \num{0.62\pm0.01}.

\section{\label{setup}Experimental Setup}

\begin{figure}
\includegraphics{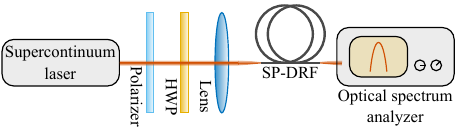}
\caption{\label{fig2}Experimental setup for studying the polarization-dependent behavior of SP-DRF.}
\end{figure}

Figure \ref{fig2} is a schematic of the experimental setup for measuring the polarization-dependent behavior of SP-DRF. A supercontinuum laser beam was passed through a linear polarizer, converting the elliptically polarized light from the source to a linearly polarized light. A broadband half-wave plate (HWP) was used to rotate the plane of polarization of the linearly polarized light. The linearly polarized light after the HWP was coupled to SP-DRF, and the output was directly analyzed by an optical spectrum analyzer to study the polarization response of the SP-DRF \cite{41,42}. The angle of rotation of the plane of polarization $\theta$ is set such that $\theta=$ \SI{0}{\degree} and \SI{180}{\degree} correspond to the minimum transmitted output power from the SP-DRF, and $\theta=$ \SI{90}{\degree} corresponds to the maximum, representing the block and pass polarizations, respectively.

\section{\label{results}Results and Discussions}

\begin{figure*}
\includegraphics{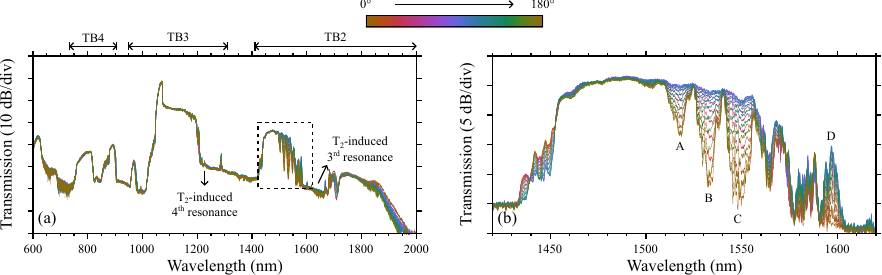}
\caption{\label{fig3}(a) Transmission spectra for varying orientations of linearly polarized light launched into a \SI{3}{\meter}-long SP-DRF. TB2, TB3, and TB4 denote the second, third, and fourth transmission bands of T\textsubscript{1}. The third resonant band of T\textsubscript{2} tubes is located in the second transmission band of T\textsubscript{1} (TB2), allowing for a high degree of polarization control. (b) Magnified plot in the wavelength range between \num{1420} and \SI{1620}{\nano\meter}.}
\end{figure*}

The transmission spectra from a \SI{3}{\meter}-long SP-DRF for varying $\theta$ from \SI{0}{\degree} to \SI{180}{\degree} in steps of \SI{10}{\degree} are plotted in Figure \ref{fig3}(a). The antiresonant and resonant bands corresponding to T\textsubscript{1} are clearly observed. The T\textsubscript{2}-induced fourth and third resonant bands are centered at \num{1266} and \SI{1638}{\nano\meter}, respectively. The resonant band at around \SI{1266}{\nano\meter} overlaps with a resonant band of T\textsubscript{1}, implying that both X- and Y-polarizations of the FM suffer high propagation loss, resulting in no polarization-dependent behavior in this wavelength range. However, the third resonant band of T\textsubscript{2} overlaps with an antiresonant band of T\textsubscript{1}. As a result, we notice strong polarization-dependent transmission caused by T\textsubscript{2} at the short wavelength edge of its resonant band centered at \SI{1638}{\nano\meter}.

We can see three distinct polarization-dependent loss peaks labeled A, B, and C in the wavelength range \SIrange[range-phrase=--]{1500}{1600}{\nano\meter} at $\theta=$ \SI{0}{\degree} and \SI{180}{\degree}, as shown in Figure \ref{fig3}(b). They correspond to the resonant coupling of the X-polarization of the FM to different dielectric cladding modes. The almost flat transmission curve for the Y-polarized beam, i.e., $\theta=$ \SI{90}{\degree}, signifies its low-loss propagation along the fiber The co-presence of resonance-induced high-loss peaks for the X-polarization and low-loss antiresonant guidance of the Y-polarization can be attributed to the resonant coupling induced by the two T\textsubscript{2} tubes in the X-direction and strong light confinement provided by the remaining tubes of the SP-DRF. There is another single polarization peak centered at \SI{1595}{\nano\meter} labeled D. This wavelength lies deep in the resonant band of T\textsubscript{2}, and hence the loss is relatively high even for the Y-polarization. Therefore, at this wavelength, the fiber can only be used as a short-length discrete polarizer. On the other hand, operations at wavelengths corresponding to A, B, and C are suitable for applications requiring long single-polarization-guiding hollow-core fiber.

\begin{figure}
\includegraphics{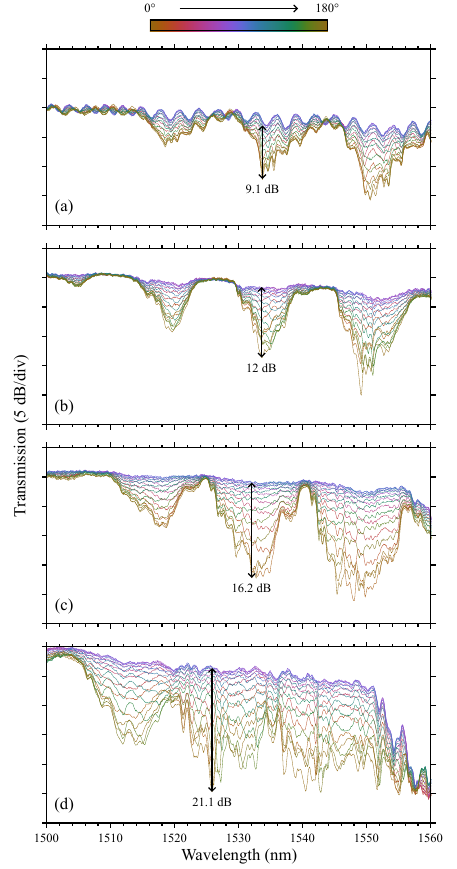}
\caption{\label{fig4}Transmission spectra for varying orientations of linearly polarized light launched into (a) \SI{1}{\meter}-long, (b) \SI{2}{\meter}-long, (c) \SI{3}{\meter}-long, and (d) \SI{10}{\meter}-long SP-DRFs.}
\end{figure}

We studied the polarization response of the SP-DRF for different fiber lengths. The transmission spectra for varying angles of input polarization are measured with different fiber lengths. We also calculated the polarization extinction ratio (PER), defined as the power ratio between the transmitted power of the pass and block polarizations for linearly polarized beams \cite{41}. The results are presented in Figure \ref{fig4}. We observe three distinct single polarization wavelengths near where A, B, and C are located in Figure \ref{fig3}(b). Since the polarization-dependent loss accumulates along the fiber, using a longer fiber generally increases the PER. For a \SI{10}{\meter}-long SP-DRF, which is the longest fiber length studied, the maximum PER is achieved around B. Tracking the peak near B at different SP-DRF lengths, we observe PERs of \SI{9.1}{\decibel} at \SI{1534}{\nano\meter} for a \SI{1}{\meter}-long fiber, \SI{12}{\decibel} at \SI{1533.5}{\nano\meter} for a \SI{2}{\meter}-long fiber, \SI{16.2}{\decibel} at \SI{1533}{\nano\meter} for a \SI{3}{\meter}-long fiber, and \SI{21.1}{\decibel} at \SI{1526}{\nano\meter} for a \SI{10}{\meter}-long fiber The bandwidth of the polarizing window with $>$\SI{12}{\decibel} PER for the \SI{10}{\meter} fiber is \SI{10}{\nano\meter} centered at \SI{1526.5}{\nano\meter}. These results constitute for the first demonstration of a polarizing AR-HCF with low-loss transmission in the pass polarization over meter-long fibers.

We observed a shift of the peak PER wavelength around B when the fiber length was changed. This is due to the variation of the loss in the pass polarization at different wavelengths. As we shall see \SI{1526}{\nano\meter} is the wavelength at which the lowest loss occurs for the pass polarization within the spectrum that we studied. Thus, the peak PER for the \SI{10}{\meter}-long fiber is observed at \SI{1526}{\nano\meter}.

Another important observation from Figure \ref{fig4} is that the PER does not increase linearly with the fiber length. That is, while the polarization-dependent loss peak deepens with the fiber length, the change is not linear. Instead, the peak becomes spectrally broadened. This is attributed to unavoidable fluctuations of $t_2$ along the fiber length, which commonly occurs during the drawing process due to minor variations in the drawing parameters, e.g., temperature and pressure. Our calculations suggest that a mere \num{\pm10} \si{\nano\meter} difference in $t_2$ can cause the third resonant band to be shifted by about \num{\pm7} \si{\nano\meter}. In fact, this acts in favor of broadening the working spectral range of the SP-DRF.

\begin{figure}
\includegraphics{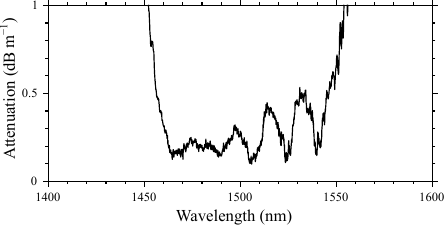}
\caption{\label{fig5}Loss spectrum calculated from the cut-back measurements.}
\end{figure}

One of the main merits of the SP-DRF is its low-loss guidance for pass polarization. Since the Y-polarization enjoys low-loss antiresonant guidance across the polarizing wavelength band, single polarization guidance can be maintained in the SP-DRF for several meters. We measured the transmission loss of the pass polarization using the cut-back method. The transmission spectra were measured first at the output of a \SI{10}{\meter}-long fiber and then at the output of a \SI{1}{\meter}-long fiber after cutting it back. The corresponding loss spectrum is shown in Figure \ref{fig5}. The loss of the pass polarization is below \SI{0.6}{\decibel\per\meter} across the entire polarizing window from \num{1500} to \SI{1554}{\nano\meter}. The loss at wavelength longer than \SI{1554}{\nano\meter} is significantly higher due to their proximity to the $t_2$-induced resonant band. Within the polarizing window, the minimum loss of \SI{0.15}{\decibel\per\meter} was measured at \SI{1526}{\nano\meter}. It is noteworthy that the loss of the pass polarization in the polarizing window is not much higher than at the wavelengths far away from the $t_2$-induced resonant band, i.e., \SIrange[range-phrase=--]{1465}{1490}{\nano\meter}, which implies excellent confinement provided by the dual cladding structure leading to low-loss antiresonant guidance of the pass polarization.

\begin{figure}
\includegraphics{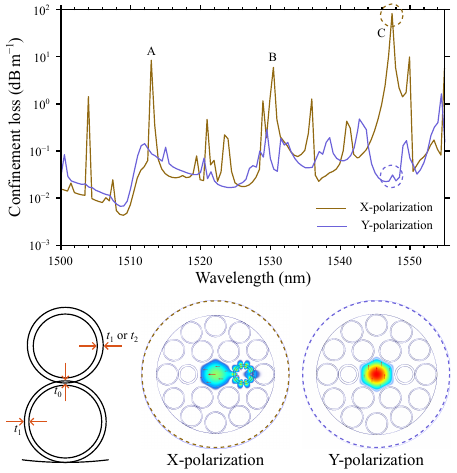}
\caption{\label{fig6}Simulated confinement losses for the X- and Y-polarizations of the FM in the SP-DRF. The simulation considers the fabricated fiber geometry shown in Figure \ref{fig1}c with $D=$ \SI{36}{\micro\meter}, $d_1/D=$ \num{0.72}, $d_2/D=$ \num{0.65}, $t_1$ \SI{1.3}{\micro\meter}, and $t_2=$ \num{2.41} and \SI{2.36}{\micro\meter} for the two T\textsubscript{2} tubes. As illustrated in the bottom-left inset, there is unavoidable overlapping of glass at the nodes between adjacent tubes in the two layers. The penetration depth $t_0/t_1$ is set to be \num{0.25}. The intensity profiles of the X- and Y-polarizations at \SI{1547.5}{\nano\meter} are shown in the bottom-right inset.}
\end{figure}

We corroborated our experimental observations against numerical simulations. Figure \ref{fig6} presents the simulated confinement loss for the X- and Y-polarizations of the FM in the SP-DRF. The parameter set considers the fabrication error associated with the overlapping of cladding tubes at the nodes between the two cladding layers. This is quantified as $t_0/t_1$, as illustrated in the bottom-left inset. We set $t_0/t_1=0.25$ based on our fabricated sample. We observe that there are three regions labeled A, B, and C, where the loss of X-polarization is much higher than that of Y-polarization. This agrees well with the three polarization-dependent transmission dips discussed in Figure \ref{fig4}. The mode profiles for the X- and Y-polarizations at \SI{1547.5}{\nano\meter} are shown in the bottom-right insets, illustrating the strong coupling between X-polarization and dielectric mode of T\textsubscript{2} tubes and excellent confinement of Y-polarization by dual-ring cladding. The simulated loss ratio between X- and Y-polarizations can reach as high as \num{2595} at \SI{1547.5}{\nano\meter}, corresponding to a PER of \SI{33.6}{\decibel} per unit length. The simulated loss of the pass polarization is significantly lower than the measurement from the fabricated fibers. This discrepancy is attributed to imperfections in the fiber fabrication, especially the discrepancy in the sizes of the two T\textsubscript{2} cladding tubes in the SP-DRF. Moreover, the wall thickness of T\textsubscript{2} cladding tubes fluctuates slightly along the fiber length. As a result of this, the polarizing window in the fabricated fiber shown in Figure \ref{fig4} is wider compared to the sharp loss peaks in the simulated spectrum in Figure \ref{fig6}. We also note that two of the cladding tubes in the SP-DRF are sticking to each other, leading to formation of nodes and additional resonances. However, these resonant wavelengths lie far outside the investigated polarizing window and hence, do not affect the results significantly. By optimizing the fabrication conditions, a perfectly uniform, non-touching cladding tube configuration can be achieved \cite{37}.

\begin{figure}
\includegraphics{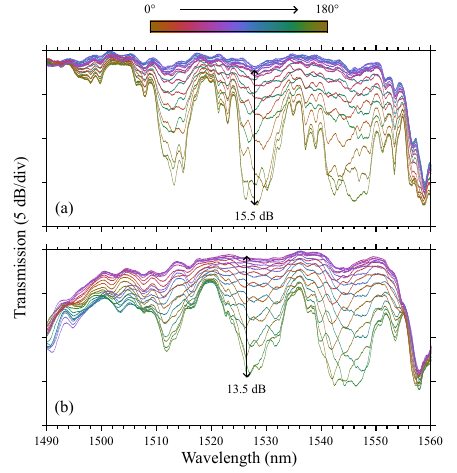}
\caption{\label{fig7}Transmission spectra for a \SI{2}{\meter}-long SP-DRF coiled into (a) two loops with a bending radius of \SI{10}{\centi\meter} and (b) four loops with a bending radius of \SI{5}{\centi\meter}, at varying input polarization angles.}
\end{figure}

For its field deployment in a compact configuration, the SP-DRF will need to be coiled. Since mechanical bending is known to induce index distortion and further the index matching-induced resonant coupling \cite{43,44}, we examined the polarization behavior of the SP-DRF under mechanical bending. The single-polarization guidance is maintained. Figures \ref{fig7}a and \ref{fig7}b show that a \SI{2}{\meter}-long SP-DRF wound into two and four loops with bending radii of \num{10} and \SI{5}{\centi\meter}, respectively, can still exhibit PER of \num{15.5} and \SI{13.55}{\decibel}. The slight improvement in performance can be attributed to the reduced degeneracy of the two orthogonal fundamental modes caused by bending \cite{45}.

\begin{figure}
\includegraphics{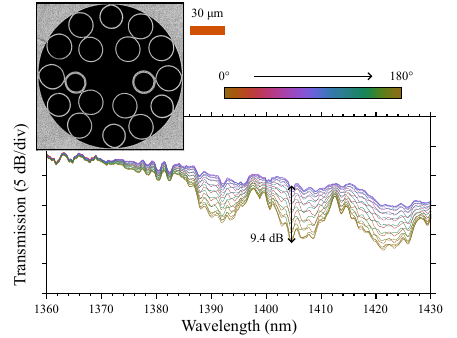}
\caption{\label{fig8}Transmission spectra for a \SI{2}{\meter}-long SP-DRF with $t_2=$ \SI{2.17}{\micro\meter} at varying input polarization angles.}
\end{figure}

The design principle of the SP-DRF is universal, and the wavelength band of single polarization guidance can be shifted to any desirable range by scaling the geometrical parameters of the fiber cross-section. This is because the resonant coupling between the dielectric cladding modes and the single polarizations solely depends on $t_2$. To verify this, we fabricated another SP-DRF with similar parameters to that shown in Figure \ref{fig1}a, but with a thinner T\textsubscript{2} wall thickness, $t_2=$ \SI{2.17}{\micro\meter}. The transmission spectra plotted in Figure \ref{fig8} show that the single polarization band is blue-shifted by \SI{130}{\nano\meter}, confirming the generality of the fiber design strategy.

\section{\label{conclusions}Conclusions}

We proposed and experimentally demonstrated the first low-loss, single-polarization AR-HCF based on a dual-ring cladding geometry. It exhibits excellent single-polarization mode guidance in the C-band with a maximum PER of up to \SI{21.3}{\decibel} and a low propagation loss of \SI{0.15}{\decibel\per\meter} for the pass polarization. The single polarization guidance and low pass polarization propagation loss can simultaneously be achieved due to the strong confinement provided by the outer-layer-cladding tubes in the dual-ring geometry. The design principle is universal, and the wavelength of operation can be engineered by simply altering the cladding tube wall thicknesses. The single-polarization guidance of SP-DRF is found to be robust against stresses caused by mechanical bending.

\section*{Acknowledgments}

The authors gratefully acknowledge the support of the Ministry of Education - Singapore under grant number MOE-T2EP50122-0019.

\section*{Author Declarations}

The authors have no conflicts to disclose.

\section*{Data Availability Statement}

The data that support the findings of this study are available from the corresponding author upon reasonable request.

\section*{References}

\bibliography{references}

\end{document}